\documentclass[%
 reprint,
 amsmath,amssymb,
 aps,
]{revtex4-2}
\usepackage{hyperref}                         % %
\usepackage[all]{hypcap}                      % %
\usepackage{graphicx}% Include figure files
\usepackage{dcolumn}% Align table columns on decimal point
\usepackage{bm}% bold math
\usepackage{xcolor}
\usepackage{caption}
\usepackage{subcaption}
\usepackage[font=small,labelfont=bf,justification=justified]{caption}
\usepackage{float}
\begin{document}

\preprint{APS/123-QED}

\title{Densities probabilities of a Bose-Fermi Mixture\\ in 1D Double well potential}% Force line breaks with \\

%\author{R. Avella}
%\author{J. J. Mendoza-Arenas}
%\author{R. Franco}
%\author{J. Silva-Valencia.}

\noindent
\author{J. Nisperuza$^1$, JP Rubio$^1$ and R. Avella $^1$}\\
\altaffiliation[]{rgavellas@libertadores.edu.co}%Lines break automatically or can be forced with \\
\affiliation{%
$^1$ Facultad de Ingenieria Aeronautica, Fundaci\'{o}n Universitaria los Libertadores, A. A. 75087 Bogot\'{a}, Colombia.\\
%$^2$ %\textbackslash\textbackslash
}%

%\author{J. J. Mendoza-Arenas}
%\affiliation{
%\affiliation{
%\author{R. Franco}
%\affiliation{%
%\author{R. Franco}
%\affiliation{%
%	Authors' institution and/or address\\
%	This line break forced with \textbackslash\textbackslash
%\author{J. Silva-Valencia}
%\affiliation{%
%	Authors' institution and/or address\\
%	This line break forced with \textbackslash\textbackslash

\date{\today}% It is always \today, today,

\begin{abstract}
We use the two mode approximation for a interacting one-dimensional spinless soft core bosons and one half spin fermions
in a double-well potential with a large central barrier. We include all the on-site boson-boson, fermion-fermion and boson-fermion repulsive contact potential represented by delta-function and considered bosonic and fermionic isotopes of ytterbium(Yb) $^{170}Yb$ and $^{171}Yb$ respectively. By means of the approximation, we find  that in the regime $U_{BF}>U_{BB}$ give rise to a immiscible phase and in the regime $U_{BB}\geq U_{BF}$ give rise to a miscible phase, that is characterized by a temporal overlap of the bosonic and fermionic probability densities. We also report that due to the Bose-Fermi interaction, the system presents an apparent destruction of the collapse-revival oscillation of boson density probability at least in the ranges investigated.

\end{abstract}

\maketitle

%\tableofcontents

\section{INTRODUCTION}
In the field of ultracold gases, there are perfectly controllable physical parameters\cite{Anglin02} and many interesting quasi-one-dimension experiments have been realized in harmonic \cite{Pitaevskii03, pethick_smith_2001}, double-well\cite{PhysRevLett.95.010402, Gati_2007}, periodic\cite{PhysRevLett.74.1542}, and bichromatic\cite{PhysRevA.80.023606} optical-lattice traps. The atoms in these experiments can be confined to different lattice sites and if two atomic species are considered, both the effective intra and inter component interactions can be tuned with great precision via the magnetic Feshbach resonance or confinement induced resonance \cite{Ospelkaus06, Hadzibabic03, Inouye04}.

Advances in laser technology allow experimental realization in one, two and three dimensions \cite{Trotzky08, PhysRevLett.98.200405, PhysRevA.76.043606} of a single double-well and an array of many copies of the double well system, know as dimers. In these systems the tunneling between the local double-well potentials is negligible, compared to tunneling inside the double well potential and each site have a well-defined and almost identical quantum state \cite{Anderlin07, Trotzky08,PhysRevA.73.033605}. These systems allow to study a very important concepts in quantum information theory, the notion of state as a linear superposition of ‘classical’ states\cite{Feynman66}, where the system can reside in a superposition of two or more degenerate states\cite{Holstein88, Chebotarev98, Garg00} and quantum tunneling, that have applications in solid-state devices, solar cells and microscopes \cite{wiesendanger_1994}.

Particular interesting phenomena that is present in systems of interacting atoms in double-well potentials are bosonic Josephson junction \cite{PhysRevA.55.4318, PhysRevLett.79.4950, PhysRevLett.95.010402}, squeezing and entanglement of matter waves \cite{PhysRevLett.98.030407,Esteve08}, matter wave interference \cite{ANDREWSC97, Schumm06}, and exact many-body quantum dynamic in one dimension. In quantum information processing, this systems has been used as a way to make quantum logic gates for ultracold neutral atoms confined in optical lattices \cite{PhysRevLett.98.070501}.

One dimensional double well potential has been used to study both theoretically \cite{PhysRevA.78.041403, PhysRevA.61.031601} and experimentally \cite{Cataliotti01, PhysRevLett.95.010402} the simplest case of Josephson effect using $^4$He and a superfluid Fermi gas with $^3$He \cite{PhysRevA.79.033627, PhysRevLett.93.120401, refId0}.

The influence of fermions onto bosons has been investigated in different mixture, for example in a mixture of $^4He-^3He$ has been reported the observation of simultaneous quantum degeneracy \cite{Pollet08, McNamara06}. In mixture of $^{87}Rb-^{40}K$ \cite{Klempt08, Karpiuk06} has been found a pronounced asymmetry between strong repulsion and strong attraction \cite{Schneider08}. Phase separation was found in a mixture of $^{41}K-^{6}Li$\cite{Lous18}. A localized phase of ultracold bosonic quantum gases of $^{87}Rb$ induced by a small contribution of $^{40}K$ fermionic atoms was observed in \cite{Wille06}. Attractive and repulsive Bose-Fermi interaction was study in a system composed by $^{170}Yb-^{173}Yb$ (attractive) and $^{174}Yb-^{173}Yb$ (Repulsive)\cite{Sugawa11}. Other Bose-Fermi mixtures that have been experimentally reported are $^7Li-^6Li$ \cite{Akdeniz02}, $^{39}K-^{40}K$ and $^{41}K -^{40}K$ \cite{Vichi98}.

Theoretically, numerous studies have been carried out to analyze the Bose-Fermi mixture (BFM). For example, the collapse in attractive mixtures has been studied numerically \cite{Roth02, Liu03, Modugno03, Adhikari04}, semi-analytically \cite{Miyakawa01, Karpiuk05}, or in the Thomas–Fermi approximation\cite{Pelster07}. The Bose-Fermi interaction induce the pairing of fermions analogous to the formation of Cooper pairs in the BCS model \cite{Bijlsma00, Heiselberg00, Viverit02} and the boson phase transition from the Mott insulator to super fluid \cite{Mering08, Bukov14, Fehrmann04}. Other studies indicate an asymmetry between the attraction and repulsion cases \cite{Best09, Albus03}, as well as phase separation, spatial modulation \cite{Polak10}, supersolid phase and charge density wave \cite{Titvinidze08}.

There is a fine line separating properties of bosons and fermions, for example, in the Bose-Fermi mapping theorem for hard core particles \cite{Fang11} realized experimentally for first time in \cite{Paredes04}, the ground-state is highly degenerate in a limit of infinitely strong boson-fermion repulsions \cite{Girardeau06} due to the freedom of fixing the sign of many-body wave function under the exchange of a boson with a fermion \cite{Guan09, Wang09, Yang09}. To find a duality relation in systems composed by spin-1/2 fermions interacting with two-component bosons \cite{Girardeau04} was used the Cheon-Shigehara mapping.

The most general one-to-one mapping between bosons and fermions in one dimension was found by M. Valiente. This general mapping not restricted to pairwise forces, is valid for arbitrary single-particle dispersion, including non-relativistic, relativistic, continuum limits of lattice hamiltonians and can also be applied to any internal structure and spin \cite{Valiente20}.

This leaves open the possibility of studying the transmutation of bosons into fermions and vice versa in a system of spinless bosons in the soft-core limit and spin one-half fermions, that is studied in this paper. In this work we focus on a repulsive interacting Bose-Fermi mixture of a few particles confined in a 1D double well potential at zero temperature.

The paper is organized as follows. The model used to describe a mixture of bosonic and fermionic atoms is introduced in Sec.\ref{sec:Bose-Fermi mixtures model}. In section \ref{sec:probabilities}, we vary the inter and intra species interactions and find two different regions and finally in Sec. \ref{sec:Conclusions} we make remarks.
%%%%%%%%%%%%%%%%%%%%%%%%%%%%%%%%%%%%%%%%%%%%%%%%%%%%%%%%

\section{Physical Model}\label{sec:Bose-Fermi mixtures model}

Due to the experimental possibility of confining quantum gases in an array of many copies of the double well system, where
the tunneling between the local double-well potentials is negligible compared to tunneling inside the double well potential, we considered the study in one of these potentials. In our study we consider the experimental setup  used in \cite{PhysRevLett.92.050405}, where the radial separation of the potential wells is $d=13\mu m$, a trap depth of $h\times4.7Khz$ and we considered the width of each well of $a=6\mu m$. This potential confines bosonic and fermionic isotopes of ytterbium(Yb) $^{170}Yb$ and $^{171}Yb$ respectively and we considered that the fermions isotopes have two internal degrees of freedom, allowing a density of fermions per site of $0\leq\rho_F\leq2$; the maximum number of scalar bosons on the ``soft-core'' boundary per site, is $n_{max}^B=2$. This value is due to the fact that in several reports has been found that the qualitative physical properties obtained for $n_{max}^B=2$ do not change when $n_{max}^B$ is increased \cite{Pai96, Rossini12}.

The general wave function in the basis of the two states that is noncommittal as to which particle is in which state and provid an accurate formulation \cite{PhysRevLett.79.4950, PhysRevA.59.620, PhysRevA.61.031601} can be expressed as
\begin{equation}
\Psi_{\pm}(x_1,x_2)=A[\psi_s^1(x_1)\psi_a^1(x_2)\pm\psi_a^1(x_1)\psi_s^1(x_2)],
\end{equation}
where $A$ is the normalization constant, plus sign is for symmetric (bosons) a minus sign is for antisymmetric (fermions) functions, $\psi_s^1(x_i)$ indicate the wave function of the ith particle in the ground state and $\psi_a^1(x_i)$ is the wave function of the ith particle in the first excited level; these wave function satisfy the orthonormal condition
\[
\int\psi_i(x)\psi_j(x)dx=\delta_{ij}.
\]
The two particles wave function at time $t$ of these vectors are
\begin{equation}
\label{twowave}
\begin{split}
&\Psi_{R,L}(x_1,x_2,t)=\\
&\frac{e^{-i\frac{E_s^1+E_a^1}{\hbar}}}{2}\Big[e^{i\Omega_1t}\psi_s^1(x_1)\psi_s^1(x_2)\pm \psi_s^1(x_1)\psi_a^1(x_2)\\
&\pm\psi_a^1(x_1)\psi_s^1(x_2)+e^{i\Omega_1t}\psi_a^1(x_1)\psi_a^1(x_2)\Big],
\end{split}
\end{equation}
where $\Omega_1=\frac{E_a^1-E_s^1}{\hbar}$ is the Bohr frequency and $\Psi_{R,L}(x_1,x_2,t)$ describe the probability of find the two particles on the right $\Psi_{R}(x_1,x_2,t)$ (plus sign) or on the left $\Psi_{L}(x_1,x_2,t)$(minus sign) of the double well.  From this we deduce the probability density
\begin{equation}
\label{denpro}
\begin{split}
&|\Psi_{R,L}(x_1,x_2,t)|^{2}=\\
&\Bigg\{\left[\frac{1}{2}\left[\psi_s^1(x_1)\right]^{2}\pm\frac{1}{2}\left[\psi_a^1(x_1)\right]^{2} \pm \cos\left[\Omega_1t\psi_s^1(x_1)\psi_a^1(x_1)\right]\right]\\
&\left[\frac{1}{2}\left[\psi_s^1(x_2)\right]^{2}\pm\frac{1}{2}\left[\psi_a^1(x_2)\right]^{2} \pm \cos\left[\Omega_1t\psi_s^1(x_2)\psi_a^1(x_2)\right]\right]\Bigg\},
\end{split}
\end{equation}
In this way the two-particle Hamiltonian for the system in one dimension is

\begin{equation}
\label{hubofear}
\begin{split}
\hat{H}_{BF}=\hat{H}_{B}^{\lambda_{B}}(x_1,x_2)+\hat{H}_{F}^{\lambda_{F}}(x_1,x_2)+\lambda_{BF}\delta(x_1,x_2),
\end{split}
\end{equation}
where
\begin{equation}
\begin{split}
&\hat{H}_{B}^{\lambda_{B}}(x_1,x_2)=\\
&-\frac{\hbar^{2}}{2m_B}\frac{d^{2}}{dx_1^{2}}+V(x_1)-\frac{\hbar^{2}}{2m_B}\frac{d^{2}}{dx_2^{2}}+V(x_2)
+\lambda_{B}\delta(x_1-x_2),
\end{split}
\end{equation}
and
\begin{equation}
\begin{split}
&\hat{H}_{F}^{\lambda_{F}}(x_1,x_2)=\\
&-\frac{\hbar^{2}}{2m_F}\frac{d^{2}}{dx_1^{2}}+V(x_1)-\frac{\hbar^{2}}{2m_F}\frac{d^{2}}{dx_2^{2}}+V(x_2)
+\lambda_{F}\delta(x_1-x_2)
\end{split}
\end{equation}
The repulsive contact potential between bosons (fermions) is represented by delta-function potential $\lambda_{B}(_F)\delta(x_1-x_2)$,  where $\lambda_{B(F)}>0$ for a repulsive interaction, $m_B(_F)$ is the mass of $^{170}Yb$ ($^{171}Yb$), $V(x)$ is the external confinement potential that is the same for both species and $\lambda_{BF}\delta(x_1,x_2)$ is the repulsive interaction between two ultracold neutral atoms of different statistic. We consider that the perturbation is sufficiently small, therefore is considered that the amplitudes of other states do not mix \cite{doi:10.1119/1.3583478}, so we use the projection of the wave function onto two states that is used in the study of BEC in a double-well potential and in the investigation of Fermi super fluid \cite{PhysRevA.77.043609}.

From the equations \ref{twowave} and \ref{denpro} we configure our system considering that the two bosons and the two fermions are on the right side. The orbital wave function of two fermions with opposite spin in a double-well potential is decomposed into a singlet and three triplet states with respect to the pseudospin defined by the double-well potential at $t=0$ and we adopt our units of length, $l=1\mu$, energy $E_a=E/\xi$ with $\xi=10^{-31}$ and time $\tau=\hbar/\xi$\cite{Avella_2016}. Henceforth, we will measure lengths, energies and time in these units.

\section{Bose-Fermi probabilities}\label{sec:probabilities}
We considerer that the initial state of our system is configured, with two spinless soft core bosons and two one half spin fermions in the right side of the double well potential. The boson-boson $\lambda_{BB}$, fermion-fermion $\lambda_{FF}$ and boson-fermion $\lambda_{BF}$ repulsive contact interaction terms, were varied.

\begin{figure}[H]
\centering
\begin{subfigure}[b]{1.0\linewidth}
\includegraphics[width=\linewidth]{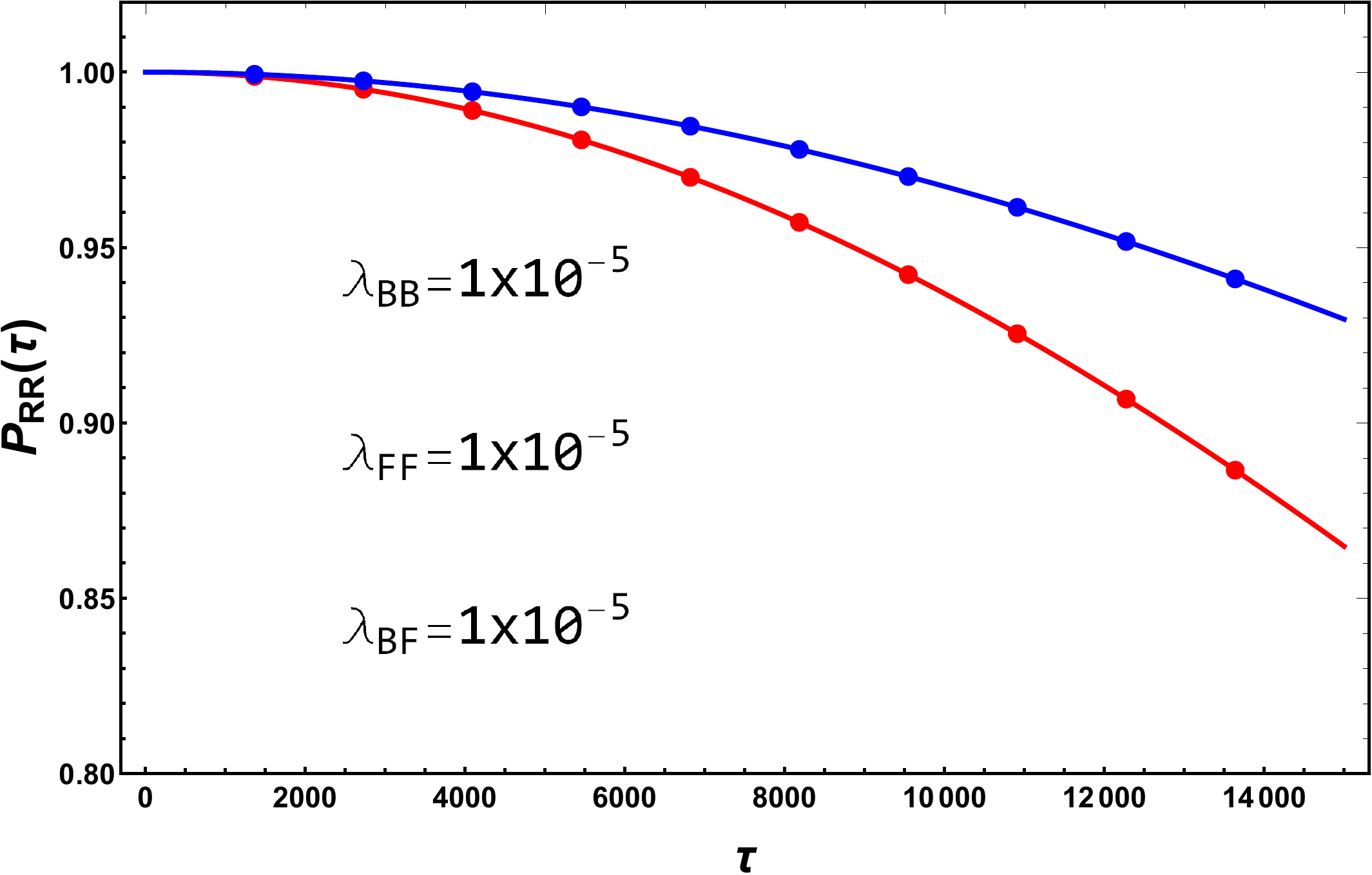}
%\captionsetup{justification=justified}
\caption{Bose (red line) and Fermi (blue line) density probabilities for: $\lambda_{BB}=\lambda_{FF}=1\times10^{-5}$ and $\lambda_{BF}=1\times10^{-5}$. The two probability densities are the same until dimensionless time $\tau\approx1363.97$.}
\label{fig:fig1}
\end{subfigure}
\begin{subfigure}[b]{1.0\linewidth}
\includegraphics[width=\linewidth]{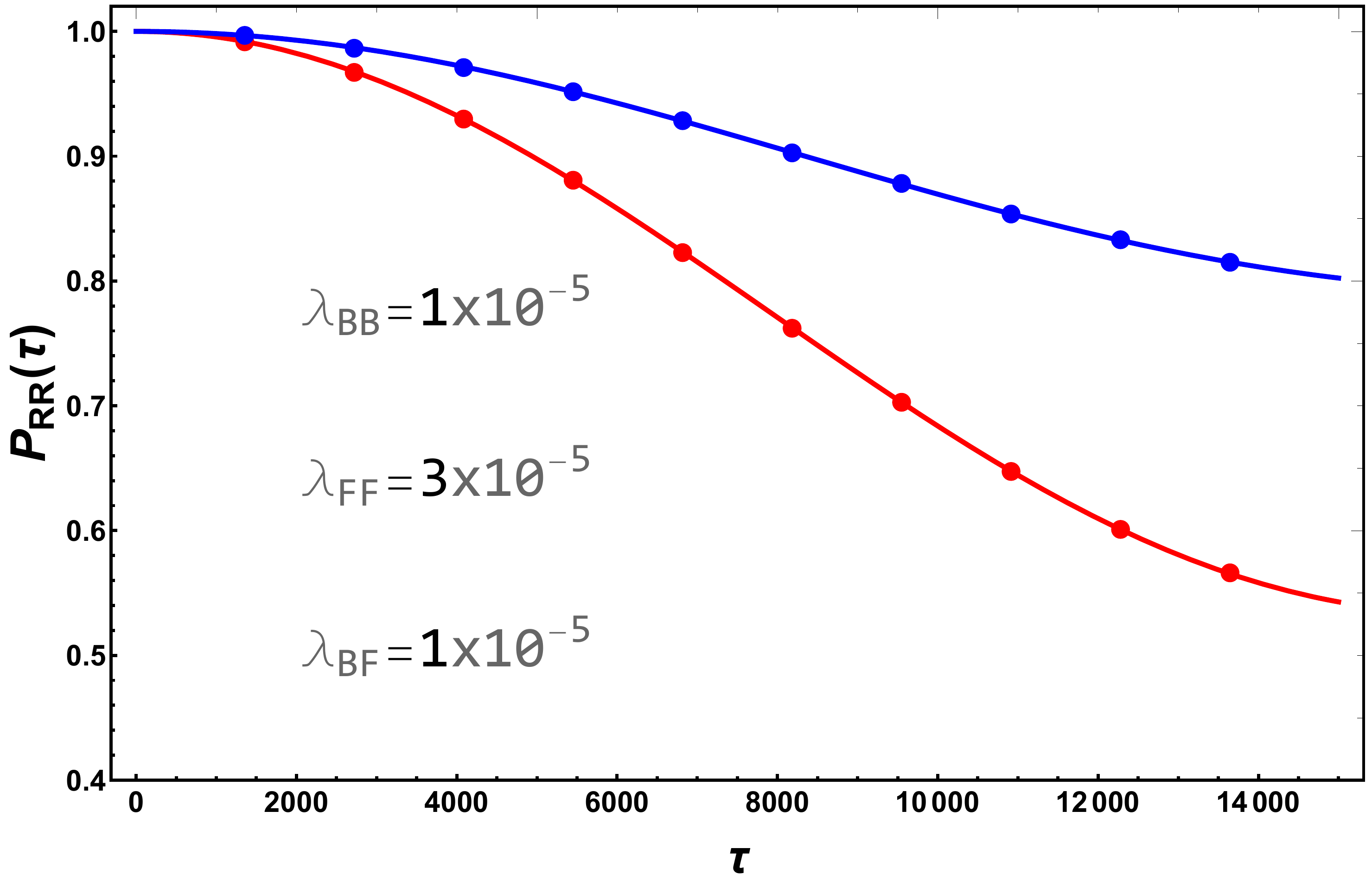}
%\captionsetup{justification=justified}
\caption{Bose (red line) and Fermi (blue line) density probabilities for: $\lambda_{BB}=1\times10^{-5}, \lambda_{FF}=3\times10^{-5}$ and $\lambda_{BF}=1\times10^{-5}$. The two probability densities are the same until dimensionless time $\tau\approx905.90$.}
\label{fig:fig3}
\end{subfigure}
\begin{subfigure}[b]{1.0\linewidth}
\includegraphics[width=\linewidth]{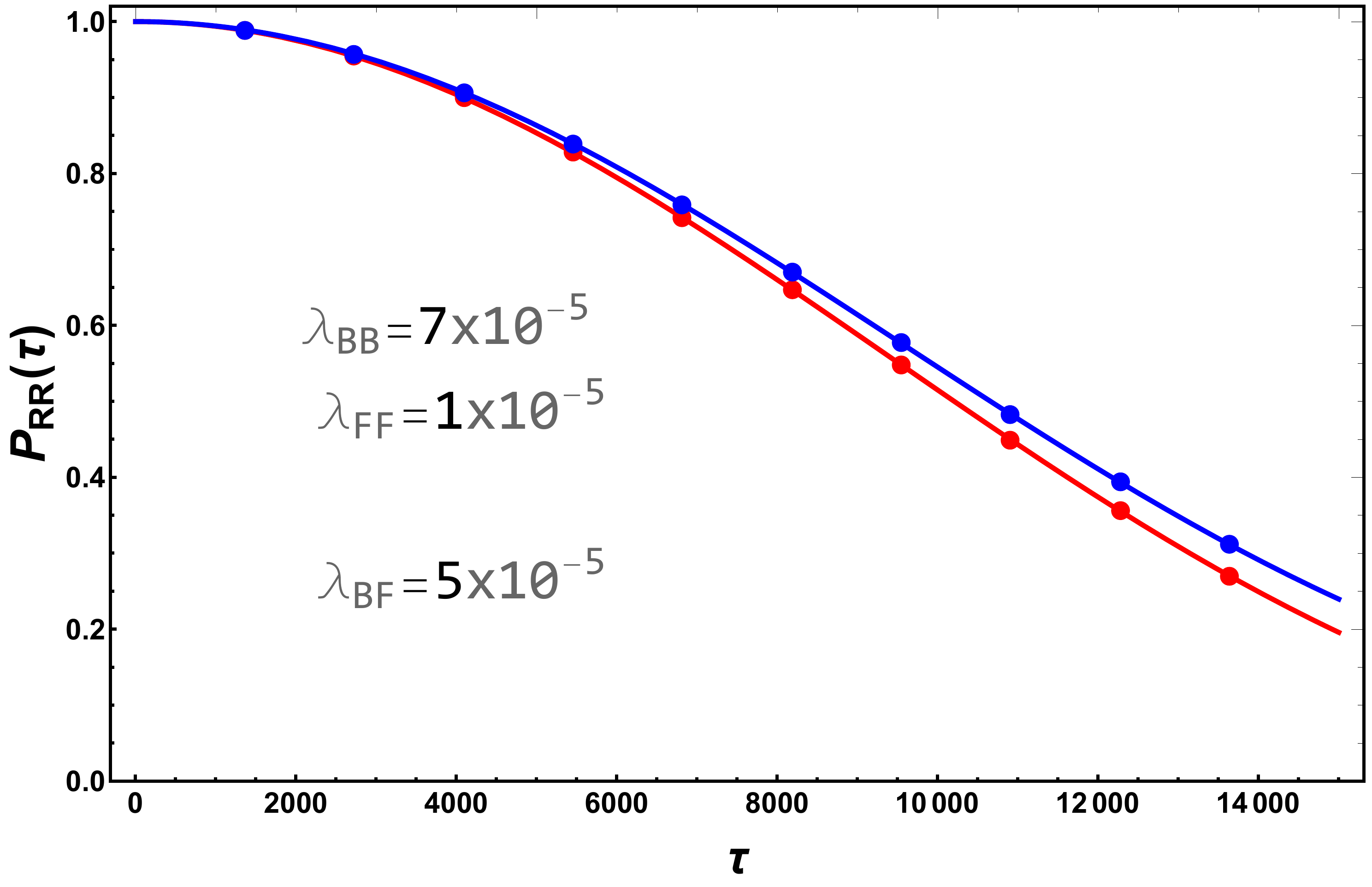}
%\captionsetup{justification=justified}
\caption{Bose (red line) and Fermi (blue line) density probabilities for: $\lambda_{BB}=7\times10^{-5}, \lambda_{FF}=1\times10^{-5}$ and $\lambda_{BF}=5\times10^{-5}$. The two probability densities are the same until dimensionless time $\tau\approx4090.97$.}
\label{fig:fig5}
\end{subfigure}
\caption{Time evolution of Bose-Fermi density probabilities on the right side of the double well, as a function of the dimensionless parameter of time $\tau$.}
\label{fig:inicial1}
\end{figure}

We found that although the initial state of the two species is considered on the right side of the double potential well, the behavior of the probability density of bosons and fermions is completely different in the regime of $\lambda_{BF}>\lambda_{BB}$, where arise a immiscible phase characterized by almost perfectly separated
fermionic and bosonic probability densities. In the regime of $\lambda_{BB}\geq\lambda_{BF}$, densities of bosons and fermions are overlapping as a function of time, as see in \ref{fig:inicial1} where $P_{RR}(\tau)$ represents the probability density of finding the two bosonic (fermionic) particles on the right side of the double well, as a function of the dimensionless parameter of time $\tau$.

Time independent miscible phase, were found theoretically in \cite{PhysRevA.97.053626}, where was studied the ground state and the nonequilibrium expansion dynamics of a Bose Fermi mixture in an one-dimensional lattice with an imposed harmonic oscillator and in \cite{nisperuza2021bosefermi}, where was studied the ground state properties of a Bose-Fermi mixture trapped in a one-dimensional optical lattice with a superimposed harmonic oscillator.

In our research we found a complete miscibility of bosons (red line) and fermions (blue line) probability densities until dimensionless time $\tau\approx1363.97$ for $\lambda_{BB}=\lambda_{FF}=1\times10^{-5}$ and $\lambda_{BF}=1\times10^{-5}$  as show in figure \ref{fig:fig1}. If fermion-fermion is greater than boson-fermion interaction, we found a miscibility of probabilities until $\tau\approx905.90$ for $\lambda_{BB}=1, \lambda_{FF}=3\times10^{-5}$ and $\lambda_{BF}=1\times10^{-5}$, which indicates that the interaction parameter $\lambda_{FF}$ induces a greater repulsive interaction in the bosons, forcing them to stay together for a shorter time and giving rise to immiscible phase more quickly, as illustrated in the figure \ref{fig:fig3}.

Is interesting to note that by increasing the value of the interaction between bosons to $\lambda_{BB}=7\times10^{-5}$ and considering that  $\lambda_{BF}>\lambda_{FF}$ the two probabilities tend to be together until $\tau\approx4090.97$ and the separation between the two probabilities tend to be closer than in the previous case. This is due to the fact that by increasing the interaction parameter between bosons, they tend to separate more quickly and this increase also affects the fermion-fermion interaction, as can be seen in figure \ref{fig:fig5}. For these values, our study revealed that due to the small interaction between fermions, they tend to stay together on the right side of the double well (blue lines) in relation to the bosons (red lines), which separate more quickly for similar values, as seen in the figure \ref{fig:inicial1}.

\begin{figure}[h!]
\centering
\begin{subfigure}[b]{1.0\linewidth}
\includegraphics[width=\linewidth]{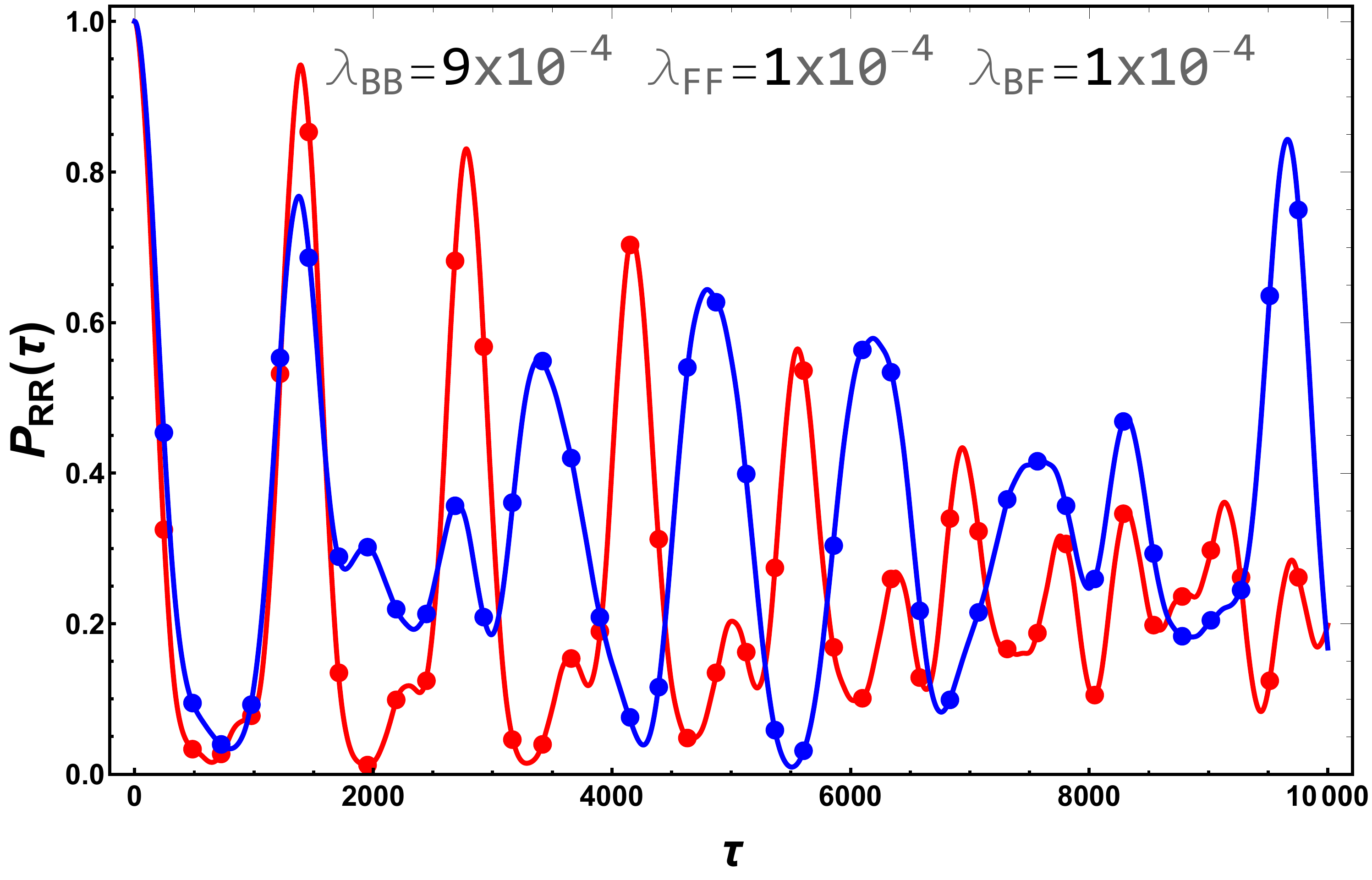}
\caption{Decrease in the boson probability amplitude (red line)}
\label{fig:fig6}
\end{subfigure}
\begin{subfigure}[b]{1.0\linewidth}
\includegraphics[width=\linewidth]{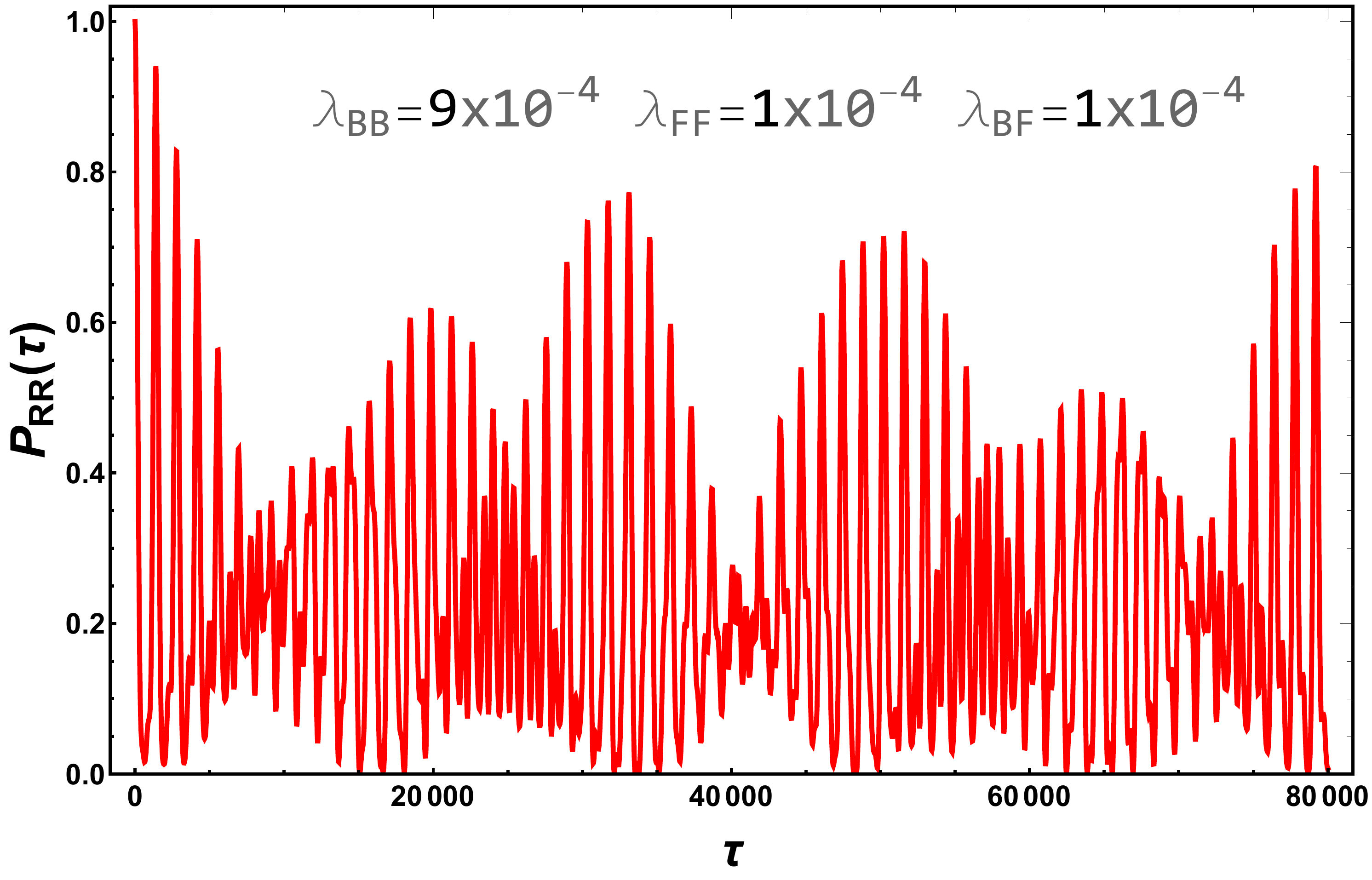}
\caption{Destruction of collapse-revival oscillation
of bosons $P_{RR}(\tau)$ due to Bose-Fermi interaction}
\label{fig:fig7}
\end{subfigure}
\caption{Apparent destruction of collapse-revival oscillation
of bosons $P_{RR}(\tau)$.}
\label{fig:westminster}
\end{figure}

By increasing the order of the interaction parameters, we found again that in the regime of $\lambda_{BB}\geq\lambda_{BF}$, densities of bosons and fermions are overlapping  when the two species begin their tunneling process on the left side of the double well in approximately the same time, again displaying a miscible phase for $\lambda_{BB}=9\times10^{-4}, \lambda_{FF}=1\times10^{-4}$ and $\lambda_{BF}=1\times10^{-4}$ that is maintained until the two species peak again on the right side of the well. The amplitud of bosons and fermions probability densities is different after the first tunneling, which is evidence of the time dependence of the miscible phase; before reaching the second tunneling, the probability of the two species ceases to be the same, giving rise to a immiscible phase as show in figure \ref{fig:fig6}. We found a transitive process from correlated Rabi oscillation of the boson particles to the apparent uncorrelated collapse-revival oscillation of $P_{RR}(\tau)$ \cite{Chang_2017, Avella_2016} due to an increasing temporary decay of the bosons oscillation amplitude (red line) as show in figure \ref{fig:fig6}. However when we increase the adimensional time, we found a apparent destruction of the uncorrelated collapse-revival oscillation as seen in the figure \ref{fig:fig7}, which indicates that the bosons tunnel through the central barrier in correlated state at least in the ranges investigated. This phenomenon indicates that due to the boson-fermion interaction, an attractive boson-boson interaction is induced and that forces the bosons to tunnel in a correlated state.

\section{SUMMARY AND CONCLUSION}\label{sec:Conclusions}

We have investigated the ground state properties of a two-component model where boson-boson, fermion-fermion and boson-fermion interaction is a weak repulsive contact potential and the two species are confined in the same one dimensional double well potential.

Tuning the ratio between the inter and intraspecies interaction strengths we found that although in the initial state both species are considered to the right side of the confinement potential, the probability densities of both species are completely different, in the regime $\lambda_{BF}>\lambda_{BB}$, giving rise to a immiscible phase. This phase occurs because fermions, despite their repulsive interaction, tend to stay longer in the initial state, while bosons tend to separate more quickly due to the fact that in addition to the repulsive contact potential between bosons, a repulsive potential is induced in them, for fermion-fermion and boson-fermion interaction

In the regime $\lambda_{BB}\geq\lambda_{BF}$ probability densities of bosons and fermions are overlapping, giving rise to a miscibility phase that disappears as a function of time. These results indicate that the immiscible and miscible phases depend of the competition between the inter and intra species interaction strength and are characterized by negligible and complete overlap of the density probabilities, respectively.

As a final attempt we increase the order of the interaction parameters, and found that the miscible phase is a time function again and as a result of the interaction between bosons and fermions the system presents an apparent destruction of the uncorrelated collapse-revival oscillation of boson density probability at least in the ranges investigated.

Our findings can help interpret experimental results in bosonic Josephson junction, squeezing, entanglement of matter waves, matter wave interference and in quantum information processing.

\section{ACKNOWLEDGMENTS}

R.A. is thankful for the support of Departamento Administrativo de Ciencia, Tecnolog\'{\i}a e Innovaci\'{o}n (COLCIENCIAS) (Grant No. FP44842-135-2017).

We gratefully acknowledge support from "Fundación Universitaria los libertadores". We are especially indebted to Mistakidis, for sending the arcticle \cite{PhysRevA.97.053626} before this publication.

%%%%% https://www.quantum-alliance.de/news-and-events/virtual-aps-march-meeting.html
%%%%% https://www.quantum-alliance.de/
%%%%%%   https://www.dfg.de/en/dfg_profile/head_office/structure/contacts/index.jsp?id=484850505356
\bibliography{apssamp}% Produces the bibliography via BibTeX.

\end{document}